# Deserialization Gadget Chains are not a Pathological Problem in Android: an In-Depth Study of Java Gadget Chains in AOSP

Bruno Kreyssig
*Umeå University*
*brunok@cs.umu.se*

Timothée Riom
*Umeå University*
*triom@cs.umu.se*

Sabine Houy
*Umeå University*
*sabineh@cs.umu.se*

Alexandre Bartel
*Umeå University*
*abartel@cs.umu.se*

Patrick McDaniel
*University of Wisconsin Madison*
*mcdaniel@cs.wisc.edu*

## Abstract

Inter-app communication is a mandatory and security-critical functionality of operating systems, such as *Android*. On the application level, *Android* implements this facility through *Intents*, which can also transfer non-primitive objects using *Java*'s *Serializable* API. However, the *Serializable* API has a long history of deserialization vulnerabilities, specifically deserialization gadget chains. Research endeavors have been heavily directed towards the detection of deserialization gadget chains on the *Java* platform. Yet, there is little knowledge about the existence of gadget chains within the *Android* platform. We aim to close this gap by searching gadget chains in the *Android SDK*, *Android*'s official development libraries, as well as frequently used third-party libraries. To handle this large dataset, we design a gadget chain detection tool optimized for soundness and efficiency. In a benchmark on the full *Ysoserial* dataset, it achieves similarly sound results to the state-of-the-art in significantly less time. Using our tool, we first show that the *Android SDK* contains almost the same *trampoline gadgets* as the *Java Class Library*. We also find that one can trigger *Java* native serialization through *Android's Parcel API*. Yet, running our tool on the *Android SDK* and 1 200 *Android* dependencies, in combination with a comprehensive sink dataset, yields no security-critical gadget chains. This result opposes the general notion of *Java* deserialization gadget chains being a widespread problem. Instead, the issue appears to be more nuanced, and we provide a perspective on where to direct further research.

## 1 Introduction

Insecure deserialization is one of the OWASP Top 10 most severe software vulnerabilities [64]. Generally, they are the result of handling untrusted data that consecutively flows into security-critical functionality. In *Java*'s Serializable API, the problem is further complicated by the presence of so-called gadget chains. That is, an attacker can craft a malicious object such that the native deserialization logic executes seemingly unrelated method calls (gadgets) leading to a security-sensitive sink. Additionally, the gadgets often stem from third-party libraries and the *Java Class Library* (JCL) itself, meaning a developer has little influence on the gadget chains present within their application.

To protect against *Java* deserialization vulnerabilities, there is a plethora of research on the detection of gadget chains within the JCL and third-party libraries [11], [14], [21], [22], [24]–[29], [32], [36], [40], [41], [43]. While this has led to advances in static program analysis and fuzzing for gadget chain detection, the artifacts are heavily biased toward the JDK and handpicked dependencies. Thus far, there is no insight into deserialization gadget chains within the *Android* platform. However, knowing that *Android*'s inter-app communication API via *Intents* employs the *Java Serializable API* makes this knowledge gap precarious. Indeed, previous work [8], [20] exploited *Intents* for a different type of attack on the deserialization routine. The attack model stays the same (see Figure 1). An attacker attempts to deliver a serialized payload from a malicious app to a more privileged app. Upon deserialization, the app is tricked into executing privileged method calls on the malicious app's behalf, i.e., a *Confused Deputy* [3], [7]. While *Peles et al.* [8] use a single serializable class with an insecure *finalize*-method, we search for sequences of method calls that ultimately trigger a security-critical sink.

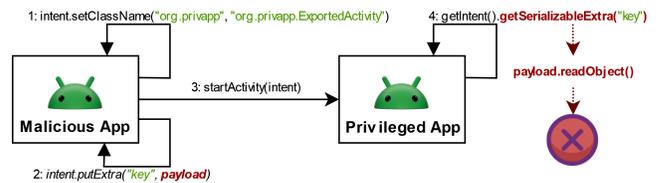

**Figure 1:** Attack Model

In this paper, we close the knowledge gap of deserialization gadget chains within *Android*. Specifically, we collect a comprehensive list of security-sensitive sink methods in *Java* and *Android* and a dataset of frequently used dependencies in *Android* applications. The latter dataset is derived from the entirety of the *Google Maven* repository, including the *Android Jetpack* suite, and through analysis of build files in *F-Droid* source code repositories. We build *AndroChain*, a static analysis tool to detect deserialization gadget chains in our dataset. We decide against using existing tools since an adequate tool for our study is required to (a) operate on the *Android SDK* instead of the JCL, (b) prioritize soundness over precision, and (c) be capable of efficiently handling a large-scale dataset. Point (b) may sound counterintuitive as recent tools [32], [36] have established the usage of fuzzers to achieve better precision. However, these works targeted a small subset of dependencies that were already known to contain gadget chains. In contrast, this paper explores new ground with the uncertainty of whether gadget chains exist to begin with. As such, we would rather filter out false positives manually through an evolving ignore list than risk missing true positives. For (a), we use the *Android* build system to generate the intermediate *Java* bytecode artifacts for *libcore* and the *Android* framework libraries preloaded in *Zygote*. Towards (c), we employ the newly released overhaul of the *Java* static analysis framework *Soot* (*SootUp* [38]), which is built for on-demand class loading. Finally, to ensure our tool performs according to the state-of-the-art, we benchmark our tool against three static gadget chain detection tools [14], [28], [40] and find that it achieves similarly sound results while significantly increasing efficiency. This step is crucial since even if we find no gadget chains, we can still confidently conclude that gadget chains are not a pathological problem in *Android*.

In a preliminary experiment, we confirm that *Android* applications widely contain insecure deserialization entry points. We also show that all but one *trampoline gadget* reachable through deserialization in the JCL are also reachable in the *Android SDK*. Further, it is possible to reach *Java* native deserialization entry points through the *Android Parcel API*. This highlights that using *Parcelable* objects for inter-app communication is not inherently safer. Overall, none of these three experiments give the impression that *Android* and its dependencies have a better security posture against insecure deserialization by design. From there, we run *AndroChain* on the collected dependency dataset, iteratively removing false positives over six executions. While the initial amount of 3 798 gadget chains appears tedious to analyze, the majority of chains connect to a much smaller number of link gadgets. Thus, with each iteration, we update an ignore list that ultimately contains 78 false positive gadgets. With this ignore list in place, *AndroChain* finds no further gadget chains.

The inability to find any security-critical deserialization gadget chains in *Android* using state-of-the-art static analysis leads us to conclude that deserialization gadget chains are not a pathological problem for *Android*. We make this statement in awareness of the limitations of *Java* static analysis (e.g., *Java Reflection* and the *Java Native Interface*). Yet, this stark difference to previous results, heavily based on *Ysoserial*, suggests that a refocus in this line of research is necessary.

Our main contributions are:

1. An analysis of insecure deserialization entry points in *Android* applications.
2. *AndroChain* - a highly efficient, open source static analysis tool for gadget chain detection based on *SootUp*.
3. A comparison of available *trampoline gadgets* in the *Java Class Library* and *Android*.
4. Concrete evidence that using *Android*'s *Parcelable* is not more secure than using *Java* native serialization.
5. The first large-scale analysis on deserialization gadget chains within *Android* and third-party libraries.

## 2 Background

### 2.1 Deserialization Gadget Chains

A deserialization gadget chain is a sequence of method calls beginning from a deserialization entry method (e.g., `readObject()`) to an attacker-desired sink method. As shown in Listing 1, the individual gadgets appear harmless in isolation. It is mandatory for a *HashTable* object to reconstruct the key hashes upon deserialization since the native hashing algorithm could be different between two *Java Virtual Machine* (JVM) instances. This implies that the `hashCode()` method of an arbitrary object is invoked (see line 6). Thereby, `hashCode()` opens up a range of polymorphic method call candidates an attacker could choose from to piece together a gadget chain.

```
1  class HashTable implements Serializable{
2    private void readObject(ObjectInputStream s) {
3      int elements = s.readInt();
4      for (; elements > 0; elements--) {
5        K key = (K)s.readObject();
6        int hash = key.hashCode();
7      }
8  }}
```

**Listing 1:** Custom Deserialization routine in HashTable.

For instance, in the *Hibernate* [37] gadget chain (see Listing 2), Java's runtime polymorphism is leveraged to invoke `TypedValue.hashCode()` from `HashTable.readObject()`. From there, by connecting gadgets from *TypedValue* (line 4) to *ComponentType* (lines 6 to 11), to *ComponentTuplizer* (lines 12 to 16), and *GetterMethodImpl* (lines 17 to 21), one can ultimately invoke an arbitrary method through `Method.invoke()` in line 22. A payload for this gadget chain is constructed by wrapping the objects in reverse order, i.e., a *HashTable* containing a key of type *TypedValue*, containing a *Type* of type *ComponentType*, etc. Note that for the chain to execute, the classes need only be present on the application's classpath (i.e., loaded in the JVM). It makes no difference whether the targeted application ever uses these classes.

```java
class TypedValue implements Serializable {
  private final Type type;
  private final Object value;
  public int hashCode() {
    //simplified, real chain via ValueHolder
    return type.getHashCode(value);
  }
}} // Type extends Serializable
class ComponentType implements Type {
  final ComponentTuplizer componentTuplizer;
  public int getHashCode(Object x) {
    componentTuplizer.getPropertyValue(x,i);
}}
class ComponentTuplizer implements Serializable {
  protected final Getter[] getters;
  public Object getPropertyValue(Object c, int i){
    return getters[i].get(c);
}}
class GetterMethodImpl implements Getter {
  private final Method getterMethod;
  public Object get(Object owner) {
    return getterMethod.invoke(owner);
}}
```

**Listing 2:** Simplified Hibernate gadget chain. [37]

This concept can easily be transferred to *Android*. The only difference is that instead of the JCL being loaded into the JVM, in *Android* the *Zygote* process loads *libcore* and the framework packages (`android.*`) and shares those with all further application processes. I.e., all gadgets in *libcore* and the framework packages are available to an attacker by default. From an analysis point of view, and since the JDK JCL and *libcore* are vastly different [30], this means we need to load the same classes into the analysis framework that are available to *Zygote*. Also, *Android* forces us to consider an additional deserialization entry method coming from the *Parcel API*. Specifically, each implementation of the `Parcelable` interface must contain an inner class of type `Parcelable.Creator`, and override the `createFromParcel(Parcel in)` method [56].

## 2.2 Gadget Chain Detection Research

The status quo in *Java* deserialization gadget chain detection has gradually evolved from the older static analysis tools *GadgetInspector* [14] and *Serianalyzer* [11] towards the incorporation of fuzzers to verify gadget chain findings (e.g., *Crystallizer* [32] and *JDD* [36]). This tendency is not surprising since all research publications on gadget chain detection have mainly focused on the *Ysoserial* [37] dataset of *Java* deserialization payloads. Within this dataset, tools are reporting extremely high quantities of gadget chains. For instance, the comparative study within the *Tabby* [28] publication shows that *Serianalyzer* reports 595 gadget chains, of which 98.6% are false positives. Hence, if applied to *Ysoserial*, it seems reasonable to optimize tools towards precision. These optimizations have, however, led to severe performance degradation in efficiency (e.g., *Crystallizer* runs for 24 hours on a single dependency) and soundness (e.g., by limiting the length of gadget chains [21], [32], [36], [47]). Some of the tools [28], [32], [36] were able to find new security-critical gadget chains on current *Java* applications. However, the scale of these real-world case studies is small (in the range of 3-6 applications/dependencies). Thus far, there is no work on detecting *Java* deserialization gadget chains on a large-scale dataset. The results of such a study would be a better predictor of to which extent gadget chains are a liability to security in current *Java* applications.

To increase efficiency, *SerHybrid* [21] introduced the notion of *trampoline gadgets* - highly polymorphic call sites that are commonly reachable in the *Java Class Library*. For instance, as shown in Listing 1, `Object.hashCode()` can be assumed as an entry point since it is known to be reachable from `HashTable.readObject()`. The idea is to then start searching for gadget chains from these *trampoline gadgets* (i.e., any class overriding `.hashCode()`) instead of from `readObject()`. Naturally, this comes at the cost of soundness. Yet, since recent publications [29], [32] still make use of this heuristic, it would be beneficial to analyze which assumed *trampoline gadgets* are actually reachable within *Android*'s *libcore* and framework SDK. It can also show that certain dependencies in the *Ysoserial* dataset would not lead to gadget chains within an *Android* application.

Research on deserialization gadget chains in *Android* is limited. We could only find a blog post [44] and a PoC [42] describing the general attack model of delivering a deserialization gadget chain with an *Intent*. Both take the *Ysoserial CommonsCollections* [37] payload as an example and do not search for new gadget chains targeting *Android*. Still, *Google* is concerned about unsafe *Intent* deserialization, listing it as a common security risk in the *Android* documentation, and strongly advises using type-safe deserialization methods that were only recently introduced in API level 33 [57].

## 2.3 Android Deserialization Entry Points

In the previous sections, we established the principle idea of deserialization gadget chains, where to find gadgets in the context of *Android*, and how to construct a payload. The final prerequisite is to find an adequate entry point within an app to transfer the payload to, thereby triggering subsequent deserialization. On a high level, *Android* facilitates communication between apps through *Intents*. An *Intent* can be delivered to an *Activity*, a *BroadcastReceiver*, or a *Service* in another app [54]. Notably, for these recipients to be reachable from outside their application context, they must also be set as exported in the application's manifest (see Listing 3, line 3). In the same scope, one also finds *intent-filter*s. To reach the *Activity* in the manifest snippet in Listing 3, the sending app would have to use one of the defined action strings (line 5). Further, an *Intent* delivers additional data to the recipient app as a primitive, a *Serializable* object, *Parcelable* objects wrapped in a *Parcel*, or a *Bundle* that acts as a container for

```
1  <activity
2    android:name="org.privapp.ExportedActivity"
3    android:exported="true">
4    <intent-filter>
5      <action android:name="org.privapp.FILTER"/>
6    </intent-filter>
7  </activity>
```

Listing 3: Exported Activity in AndroidManifest.xml

the aforementioned three. Apart from primitives, all formats evidently are deserialization entry points via the methods:

```
Intent.getSerializableExtra(String name)
Intent.getParcelableExtra(String name)
Intent.getBundle().getSerializable(String key)
Intent.getBundle().getParcelable(String key)
```

However, it does not stop here. Only recently, with *Android* API level 33 and the *LazyBundle* patch[19], was automatic un-bundling removed. This means, if an app retrieves only a primitive, e.g., with Bundle.getShort() (see Listing 4), any other (including unused) values in the *Bundle* will still be initialized and, if necessary, deserialized [8]. Furthermore, this behavior also applies to primitive *Intent* deserialization methods such as Intent.getStringExtra(), as internally these values are wrapped inside the mExtras bundle [54].

Since, at the time of writing, 47.60% of *Android* devices run on an API level below 33 [62], the mere usage of *Bundles* in inter-app communication still creates an unsafe deserialization entry point for about half of *Android* users.

```
1  class ExportedActivity extends Activity {
2    void onCreate(Bundle savedInstanceState) {
3      // setContentView, restore state, ...
4      Intent i = getIntent();
5      S s = (S) i.getSerializableExtra("key");
6      P p = (P) i.getParcelableExtra("key");
7      i.getBundleExtra("bundle").getShort("key");
8  }}
```

Listing 4: Example of an *Android* unsafe deserialization entry point

Given an app with an exported *Activity* as described in Listings 3 and 4, a malicious app triggers a deserialization gadget chain by sending a crafted intent as shown in Listing 5. Bear in mind that the payload object in line 4 needn't be of the expected cast type in Listing 4 since the deserialization logic and, thus, the gadget chain is executed before the cast is resolved.

```
1  Intent i = new Intent("org.privapp.FILTER");
2  i.setClassName("org.privapp"
3    ,"org.privapp.ExportedActivity");
4  i.putSerializable("key", payload);
5  startActivity(i);
```

Listing 5: Delivering a deserialization payload via an *Intent*

The end goal of arbitrary method invocation in another app is to abuse that app's higher permission set. This is commonly known as a *Confused Deputy* attack (see Figure 1).

### 2.4 Motivation

Before proceeding with our core research questions on detecting *Java* deserialization gadget chains in *Android*, we verify the presence of insecure deserialization entry points in apps as described in theory above. Given that we aim to achieve a *Confused Deputy* exploitation, we are interested in a dataset of apps that are likely to be installed on an *Android* device. We build all *Android* default applications[1] (e.g., Settings, Calendar, and Contacts) directly from AOSP. Furthermore, we download the top 100 most installed apps [60] from *AndroZoo* [9], [34]. Then, we implement a lightweight static analysis with *Androguard* [35], which probes a given APK file for insecure deserialization entry points by (1) extracting all exported *Activities*, *Services*, and *BroadcastReceivers*, and then (2) searching for *Intent* deserialization methods invoked in the body of the respective recipient methods: onCreate, onReceive, or onStartCommand. Note that this analysis is flow-insensitive, which means a call to a method of *Bundle* may well refer to an app's savedInstanceState *Bundle* and not the *Bundle* received from an *Intent*. Therefore, in Table 1 we limit ourselves to the deserialization entry points from *Intents*: getSerializableExtra(), getParcelableExtra(), and also all primitive deserialization methods, for API levels < 33.

|  | **AOSP** (49) | **Popular Apps** (100) |
| --- | --- | --- |
| Serializable | 0 | 9 |
| Parcelable | 8 | 46 |
| *API < 33* | 22 | 88 |

Table 1: Exported deserialization entry points. For API levels < 33 all Intent.get() methods are also considered an entry point.

The results of our preliminary experiment are presented in Table 1. Indeed, it is possible to find deserialization entry points regardless of API level and underlying serialization API (*Parcel API* or *Java native Serialization API*). Even though AOSP apps make no use of getSerializableExtra(), 22 apps still expose a Serializable entry point in API levels < 33 (about half of *Android* devices [62]). One of these is the *Settings* app via the exported WifiDialogActivity [58] (see Listing 6). This is quite precarious since the *Settings* app is the ideal target for a *Confused Deputy*. It is granted 138 permissions, among others WRITE_SECURE_SETTINGS.

In widely installed apps outside of AOSP, the amount of deserialization points increases significantly. Listing 7 shows one of the insecure deserialization entry points we found from

---
[1]Apps that reside in the packages/apps directory of AOSP [55]

```
1  public class WifiDialogActivity {
2    public static final String KEY_CHOSEN_WIFIENTRY_KEY
3      = "key_chosen_wifientry_key";
4    public void onCreate(Bundle bundle) {
5      mIntent = getIntent();
6      // lines 122-131 omitted
7      mIsWifiTrackerLib = !TextUtils.isEmpty(mIntent
8        .getStringExtra(KEY_CHOSEN_WIFIENTRY_KEY));
9  }}}
```

**Listing 6:** `WifiDialogActivity` in *Android Settings* app. Arbitrary deserialization is possible in API levels < 33 through `Intent.getStringExtra()` (line 8).

our analysis. It becomes clear that widely distributed apps like *Google Photos* continue to rely on the type-unsafe deserialization methods (lines 6, 10). Moreover, given a deserialization gadget chain, finding an adequate entry point is as trivial as described in the previously used toy example (see Listing 4). Together with the *Settings app*, it proves that research on deserialization gadget chains as an attack vector is imperative.

```
1  class GuidedThingsConfirmationActivity {
2    public final void onCreate(Bundle bundle) {
3      // omitted decompiled jadx code
4      if (getIntent().hasExtra("explore_type")) {
5        alijVar = (alij) getIntent()
6          .getSerializableExtra("explore_type"); }
7      // omitted decompiled jadx code
8      if (getIntent().hasExtra("cluster_type")) {
9        akprVar = (akpr) getIntent().
10         .getSerializableExtra("cluster_type"); }
11 }}
```

**Listing 7:** Deserialization entry point decompiled from *Google Photos* APK, in `GuidedThingsConfirmationActivity`

## 3 Methodology

To analyze whether deserialization gadget chains exist within the *Android* ecosystem, we build *AndroChain*, a tool that is both capable of handling a large dataset of *Java* bytecode and still produces meaningful results according to state-of-the-art *Java* deserialization gadget chain detection tools.

### 3.1 Tool Design

At its core *AndroChain* perpetually executes the loop in Figure 2 until it reaches a certain depth or until the work list is empty. Note that in doing so, *AndroChain* is not required to maintain a full call graph in memory, which for a large dataset would be overly expensive [33]. Gadgets are implemented as a simple linked list, which can be, for the purpose of output, backward-reconstructed from sink gadget to source. This reconstruction takes place as part of the **Apply Filters** stage, in which the current gadget's method signature is compared with a list of sink methods. The filters further discard:

1. **native methods** since we only analyze *Java* bytecode.

2. **non-static methods in non-serializable classes** since implementing the *Serializable* interface is mandatory for objects to be (de-)serialized.

3. methods in an **ignore list** of known false positives.

4. methods which were **already visited**.

Towards (4), the tool keeps track of the method signatures of each examined `InvokeExpr`. Doing so significantly increases performance, especially for virtual invocations with many call candidates. It does, however, come at the cost of some soundness. For example, given an entry point gadget $E$, link gadgets $A$, $B$, and $C$; and sink gadget $S$, forming the gadget chains $E \rightarrow A \rightarrow B \rightarrow S$ and $E \rightarrow C \rightarrow B \rightarrow S$: the tool will only output either of the two chains since it will determine that gadget $B$ was already visited when computing the latter chain. However, this still means a gadget chain leading to $S$ via $B$ is found. In case the call $A \rightarrow B$ turns out to be a false positive, we can add $A$ (or the call edge) to the ignore list. When we rerun the tool, we now find the chain $E \rightarrow C \rightarrow B \rightarrow S$. Recall that in this work our objective is to find **whether any deserialization gadget chains exist in the *Android* ecosystem** to begin with. This research question requires a different tool than all previous publications that researched **how many gadget chains exist in *Ysoserial* libraries**.

Returning to Figure 2, those gadgets that were not filtered out will proceed to intraprocedural analysis. We use type propagation to precisely determine polymorphic call sites in scenarios like Listing 8. Specifically, we store the types a gadget was invoked with and implement a `ForwardFlowAnalysis` to propagate method parameter types within the gadget method's body to the polymorphic call sites.

```
1  class Gadget implements Serializable {
2      String s;
3      public static void foo(Object o) {
4          o.toString(); }
5      void readObject(ObjectInputStream s) {
6          s.defaultReadObject();
7          Gadget.foo(s); }}
```

**Listing 8:** Type propagation: with `Gadget.foo()` (line 7), the parameter type (String) needs to be propagated to the next call site. Otherwise, we receive false positives of the kind `Gadget.foo()` → `Object.toString()` instead of `Gadget.foo()` → `String.toString()`.

We also implement a lightweight taint analysis using heuristics such as field insensitivity and assuming the return value of a tainted method call is also tainted. For our purpose, this is sufficient since (1) doing so leans towards soundness and (2) we do not discard gadget chain findings based on taint analysis. We only output the taint values of a gadget chain for convenience during manual analysis.

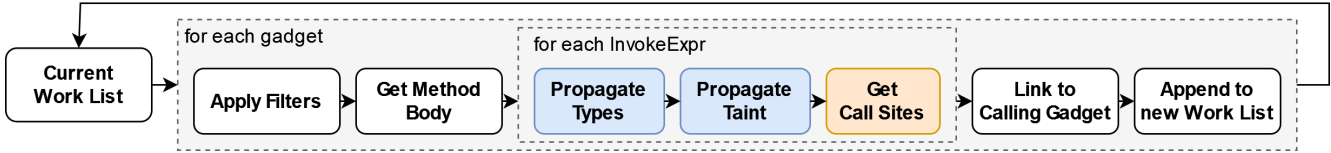

**Figure 2:** Gadget Chain Detection Loop. ▨ - Intraprocedural analysis components ▨ - Interprocedural analysis.

In most cases, we use class hierarchy analysis (CHA) [1] to resolve polymorphic call sites. Therefore, we explore the class hierarchy to add all reachable method implementations: from the declared class type itself or its first parent class implementing the method down to all its child classes. E.g., in Listing 9, `b.foo()` (line 11) could, if the property b (line 9) has a runtime type of C, invoke `C.foo()`. It could also, if b has the runtime type B, invoke the inherited `A.foo()` from its superclass A. In this case, one must also keep track of the caller being of type B so that `this.bar()` (line 2) connects to `B.bar()`. However, when new objects are instantiated during the deserialization process, we more accurately resolve virtual calls with variable type analysis (VTA) [2]. In these cases, we forgo the filter checking that the callee's class is serializable.

```java
class A {
    public void foo() { this.bar(); }
    public void bar() {}}
class B extends A implements Serializable {
    public void bar() { sensitive(); }}
class C extends B {
    public void foo() { sensitive(); }}
class Gadget implements Serializable {
    B b;
    void readObject(ObjectInputStream s) {
        b.foo(); }}
```

**Listing 9:** Class hierarchy analysis example. The call `b.foo()` in line 11 can connect to the subclass `C.foo()` or superclass `A.foo()`.

Bringing it all together, we have a lightweight static gadget chain detection tool. With our filter for only visiting call sites once, we avoid path explosion at the cost of not finding all gadget chains. This is only a limitation if *AndroChain* is run once. Updating an ignore list ensures that the remaining gadget chains are found in further executions. Still, our tool is subject to the limitations of *Java* static analysis. That is, dynamic language features such as *JNI* and *Java Reflection* [13], [15]. However, these limitations also apply to all other available gadget chain detection tools [47]. In fact, we have more visibility of reflective method calls through not culling untainted call edges (i.e., we see an untainted `Method.invoke()` sink and can then manually investigate where it leads to). Additionally, later in Section 4, we also run *AndroChain* with serializable subtypes of `InvocationHandler.invoke()` as an entry point, which gives visibility over dynamic proxies.

### 3.2 Tool Evaluation

The evaluation is run on a Dell Latitude 7440 laptop, with a 10-core Intel Core i5-1345U processor (1.60 GHz) and 32 GB of RAM. *AndroChain*, is written on top of the *SootUp* [38] (*v1.3.0*) *Java* static analysis framework in 1 745 LoC. We share *AndroChain*'s source code in accordance with the *Open Science Policy* (see Section 8).

We compare our tool to three purely static gadget chain detectors: *GadgetInspector* [14], *Tabby* [28], and *SerdeSniffer* [40]. We choose these tools because of their availability and soundness [47]. In contrast, hybrid tools like *Crystallizer* [32], *ODDFuzz* [26], and *JDD* [36] trade soundness for higher precision by employing a guided fuzzer to construct and test payload objects. We benchmark both against *Ysoserial* [37] and the synthetic *Gleipner* benchmark [47]. The latter was only released a week prior to submission which is why we describe *Ysoserial* in detail and refer to Appendix B for a summary of running the *Gleipner* benchmark.

We merge all 41 dependencies used in the *Ysoserial* repository into a single JAR file. On the one hand, this benchmark is sizable. It comprises 54 302 classes and 457 601 methods in combination with the JDK. On the other hand, it contains at least 34 confirmed gadget chains [37]. All three tools are, by design, configured to use different sink method lists and operate from a different set of source methods. For a meaningful comparison, we run our tool with the same source and sink methods, which is why, in the final three rows of Table 2, we get distinct results. Furthermore, we limit the maximum gadget chain search length to 8 when testing the benchmark on *Tabby*, as this tools execution time increases exponentially over the gadget length [2].

Table 2 shows the results of running the tools on the *Ysoserial* benchmark. We consider a gadget chain to be found by both tools if the gadgets invoking the sink method match. This ensures that gadget chains with slightly different routes (e.g., to `Object.hashCode()`) that, however, are later identical will still be counted as the same gadget chain. Doing so is in alignment with our aim of finding whether any gadget chains exist within the *Android* platform. Also, Table 2 makes no claim about the precision of our tool compared to the other three. Our tool produces many more false positives which we alleviate through the usage of an ignore list. Maintaining and iteratively updating an ignore list enables *AndroChain* to maximize both on efficiency and soundness.

---
[2]The tool *tabby-vul-finder* released by the same authors as *Tabby* also uses a maximum depth of 8 [45].

|  | **Gadget-Inspector** | **Tabby** | **Serde-Sniffer** |
|---|---|---|---|
| *length* | ∞ | 8 | ∞ |
| *source* | all | readObject | readObject |
| *total* | 71 | 441 | 138 |
| ∑ *unique* | 4 | 45 | - |
| *time* | 2m53.25s | 18m37.633s | 2h5m26s |
| | *AndroChain* | | |
| *total* | 169 | 170 | 83 |
| ∑ *unique* | 120 | 127 | - |
| *time* | 41.36s | 41.67s | 39.18s |

Table 2: Comparison of *AndroChain* (AC) to other Gadget Chain detectors. A column describes the configuration (top section), results of the tool benchmarked against (mid section), and *AndroChain*'s results using that configuration (bottom section).

We now investigate gadget chains uniquely detected by the other three tools. *GadgetInspector* output two gadget chains using implementations of *Groovy Closures* as a source method - an entry point irrelevant to the analysis of *Android*. Then, two other chains are the result of a bug in the tool, where it not only extracts the *Java Class Library* from the JVM classpath but also any other dependencies loaded in the JVM while running the tool. This leads to gadgets polluting the dataset that weren't part of it to begin with. Among the 45 unique gadget chains *Tabby* finds in *Ysoserial*, 41 are the result of lacking type propagation. Specifically, when executing Java's `AccessController.doPrivileged()` with anonymous classes extending *PrivilegedAction* or *PrivilegedActionException*, *Tabby* sees these objects as being of any subtype of the respective base class in later gadgets. One other false positive occurs due to the used *Cypher* query (see Listing 10). It so happens that *Apache Wicket's* `File` class [46] is serializable and has a `readObject()` method, which does, however, not match the signature of the default *Java* deserialization method `readObject(ObjectInputStream)`. The *Cypher* query still picks this up as an entry point since the method name and *Serializable* interface match.

```
MATCH (e:Method {NAME:"readObject", IS_SERIALIZABLE:true})
MATCH (s:Method {IS_SINK:true})
CALL tabby.beta.findJavaGadget(s, "<", e, 8, true)
YIELD path WHERE NONE(n IN nodes(path) WHERE
n.CLASSNAME IN ["java.io.ObjectInputStream"]) RETURN path
```

Listing 10: Tabby path-finder query

Two unique detections by *Tabby* referred to the same gadget chain, only that the tool had merged two method call edges into one. Finally, two gadget chains in the *C3P0* dependency were not found due to *AndroChain*'s method cache ensuring gadgets are only visited once. For both of these chains, we add the gadget of the initial finding to the ignore list and rerun our tool, whereupon it finds the two remaining gadget chains. This proves *AndroChain*'s ignore list as an adequate strategy to soundly detecting gadget chains.

The *SerdeSniffer* tool runs through three analysis stages: a pre-processing stage using *Soot*, followed by static analysis with custom *Doop Datalog* rules, and a final verification stage in a *Neo4J* database [40]. Unfortunately, the queries used in the final stage are not provided with the tool artifact. We contacted the authors for clarification but received no response. Therefore, we are left with an intermediary output from *Doop* displaying gadget chains only by their source and sink method. Without the remainder of the detected gadget chain, it is not feasible to compare the findings with *AndroChain*. However, this does not prevent making directionally correct observations. Due to its incompleteness, *SerdeSniffer* could not have been employed for our analysis on *Android*. Moreover, the time consumption of the first two analysis stages is significantly higher than for *AndroChain*, excluding any graph search used in the third stage. *SerdeSniffer* is geared towards analyzing the *JDK* with few configuration options outside of declaring the JAR files to analyze. Finally, even though the amount of found gadget chains is higher compared to our tool, it is likely that those are variations leading to the same sink gadget, as is apparent in comparison with *Tabby* in Table 2.

Overall, from the comparison with the three static analysis tools, we can confidently say that in terms of soundness, *AndroChain* performs according to the state-of-the-art. Our tool finds all true positive gadget chains found by *Tabby* in *Ysoserial* and is comparably sound on *Gleipner* (see Appendix B). *Tabby* can undeniably be considered state-of-the-art due to its recency (2023), popularity (1 300 *GitHub* star rating), and usage in further publications: *GCMiner* [27] and *ODDFuzz* [26]. Moreover, *AndroChain* is not constrained in performance by gadget chain length and is capable of processing a large-scale dataset. Consequently, our requirements to enable a meaningful analysis of the *Android* framework and libraries for app development are satisfied.

## 4 Investigating Android for Gadget Chains

In this section, we first outline the acquisition of a comprehensive *Android* dependency dataset and security-critical sink method list. Then, by showing that the *Android* framework SDK and the JCL contain the same *trampoline gadgets* (**RQ1**) and the *Parcel API* being susceptible to insecure deserialization (**RQ2**), we show that *Android* is not more robust against deserialization gadget chains by design. Finally, we analyze the entire dataset for gadget chains (**RQ3**).

We use the same hardware setup as outlined in Section 3.2. All intermediate AOSP build artifacts were built from the `android15-release` branch with the `aosp_arm64-trunk_staging-user` build target. The data collection and analysis pipelines were implemented in *Python*. For reproducibility, we release the build artifacts and *Python* scripts in Section 8.

## 4.1 Data Collection

### 4.1.1 Android Java Bytecode dataset

We consider three data sources:

1. Intermediary build artifacts from the *Android* OS source.
2. Dependencies from the *Google Maven* repository.
3. Other dependencies frequently used in *Android* apps.

As stated in Section 2, the analysis domain must contain *Android*'s core libraries loaded into *Zygote*. *Libcore* and the base framework cannot be trivially extracted from the *Android* image since they are precompiled in ART (*Android Runtime*) machine code. However, during the *Android* build process, intermediary *Java* bytecode artifacts are placed in the `out/target/common/obj/JAVA_Libraries` directory before they are optimized to ART. Furthermore, with the `mmm` build command [52], we can restrict the build process to only build the libraries we are interested in: `libcore` and `frameworks/base`.

The other two mentioned data sources cover common dependencies used in *Android* applications. We assemble two pipelines to collect the JAR file artifacts from those (see Figure 3). The *Google Maven* repository (center of the figure) contains official libraries, most notably the *Android Jetpack* suite, to support app development. We assume, per default, that any dependency in this repository could supply gadgets for a deserialization gadget chain within an *Android* application. We automate the download process of dependencies from the Google Maven Repository web listing [59]. To do so, we analyze the requests for loading the repository web page. The listing is constructed through a master index file that further references group index files containing the dependency names and their respective versions (see Listing 11). With the information from the group listings, one can construct the download URL as shown in lines 4 and 5.

```
1  dl.google.com/android/maven2/master-index.xml
2  dl.google.com/android/maven2/<package>/group-index.xml
3
4  dl.google.com/android/maven2/<group>/<dependency_name>/
5    <version>/<dependency_name>-<version>.jar
```

**Listing 11:** Google Maven repository listing.

For each dependency, we download either the latest non-alpha or if only alpha releases are available, the latest alpha version. If needed, we extract the *Java* bytecode *classes.jar* from dependencies provided in AAR format.

Where the *Google Maven* repository is an excellent source for official *Android* app dependencies, *F-Droid* [49] is a good indicator for all other third-party libraries used in *Android* app development. Specifically, we can retrieve the URL references to the source code repositories from the *F-Droid* app catalog (Figure, 3 top left) and continue to download the build files. From these, we then extract the dependency names. For automation, from the 3 703 repositories listed on *F-Droid*, we consider all repositories hosted on either *GitHub* (2 937) or *Gitlab* (374), accounting for 89.41% of the entire listing.

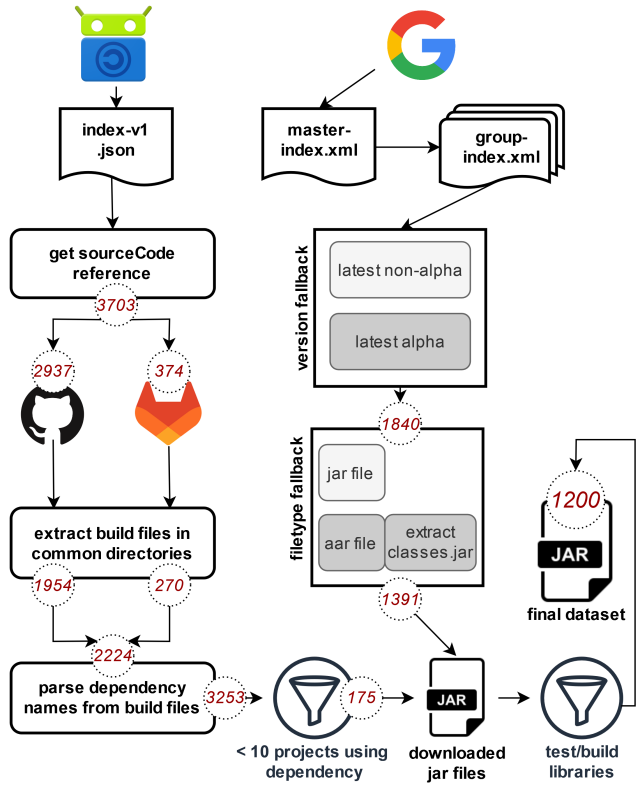

**Figure 3:** Dependency dataset collection pipeline, including number of build files/dependencies at each stage.

We leverage the default project structure used in *Android* application development [48] to find build files in the respective source code repositories. Specifically, we query the *GitHub* and *GitLab* APIs for any matching build files in Listing 12. Note that we exclude the `build.gradle` file at the project root since it is only used for plugin declarations. Dependencies, on the other hand, reside in the `app/` or `android/app` (for multiplatform projects) subdirectories. In addition, a version catalog (`libs.versions.toml`, line 5) may be used to define dependency variables that are referenced in *Gradle* build files. In this case, we can extract dependency package name literals directly from the catalog. By matching the path strings in Listing 12, we retrieve build files for 1 954 (66.5%) *GitHub* and 270 (72.19%) *GitLab* repositories referred by *F-Droid*.

From those build files, we parse 3 253 distinct dependencies, of which 175 occur 10 or more times across the projects,

```
1  app/build.gradle
2  app/build.gradle.kts
3  android/app/build.gradle
4  android/app/build.gradle.kts
5  gradle/lib.versions.toml
```

**Listing 12:** Source code repository dependency files

and download those manually in their latest version. Finally, from all downloaded JAR files, we remove test/mock dependencies, JVM-stub libraries, and *Gradle* plugins, all of which should not be included as runtime dependencies and thus cannot provide gadgets. This leaves us with **1 200** JAR files.

#### 4.1.2 Android Security Sensitive Sinks

Since *Java* sink methods (e.g., `Method.invoke()`) for the most part also exist in *Android*'s *libcore*, we port those lists from existing gadget chain detection tools [11], [14], [28], [36], [40] (links in Appendix A).

*SuSi* is a machine-learning tool to identify *Android-specific Java* sink methods [6]. We initially planned to rerun *SuSi* on a recent *Android* framework JAR. However, we were not able to build the tool since it relies on an old version of *FlowDroid* [4], for which we could neither find the source code nor release artifact. Therefore, we opt for a middle ground by including the historic sink list for *Android* 4.2 included in the *SuSi* repository while also detecting sink methods through the `@RequiresPermission` annotation from the *Android SDK* source code. As such, our final sink method dataset contains 4 616 method signatures. Note that since the *SuSi* sink list is generated through machine learning and with an old API, we might have to remove sink methods during experimentation.

| | Java Sinks | | | | Android sinks | |
|---|---|---|---|---|---|---|
| Serian alyzer | GI | Tabby | JDD | Serde-Sniffer | SuSi 4.2 | Source |
| 9 | 37 | 62 | 709 | 11 | 2 948 | 932 |
| | | 738 | | | | 3 878 |

**Table 3:** Sink dataset. The numbers represent the amount of sink methods found when loading the dataset into a SootUp view.

### RQ1: Which *trampoline gadgets* are reachable using the core SDK alone?

*Trampoline gadgets* (see Section 2.2) are a widely used heuristic in gadget chain detection since they are frequently part of known gadget chains in the *Ysoserial* repository. From *Ysoserial*, we find 6 *trampoline gadgets*, accounting for 22 out of the 34 listed gadget chains. Further, *Cao et al.* [29] list `compareTo` as a trampoline, which is why it is also included in Table 4. To compare reachability between the JDK and *Android*, we run our tool on the JCL and *Android SDK* with the 7 *trampoline gadgets* set as sink methods and `readObject()` and `readResolve()` as entry points. The results in Table 4 show that only the `Object.toString()` trampoline is available in the JCL but not from classes loaded in *Zygote*. The discrepancy arises due to the JDK's `BadAttributeValueException` not being part of *libcore*. It should, however, be noted that this gadget is only available up to JDK version 14 [18]. None of this indicates that *Android* is less susceptible to gadget chains than the JDK.

We do not pose the same question for the *Parcel API* entry point, `createFromParcel()`, since the deserialization of *Parcelable*'s is self-contained. This means all properties of a *Parcelable* are explicitly reconstructed. To find out whether the *Parcel API* is a safe alternative to native deserialization, we investigate further in the next research question.

### RQ2: Is it possible to trigger *Java* native deserialization through *Parcelable* entry points?

Knowing that `Parcelable.Creator.createFromParcel()` takes a *Parcel* as an input parameter and, hence, could reconstruct fields with `Parcel.readSerializable()`, it seems feasible that a *Serializable* entry point could be reachable through deserialization of a *Parcelable*. If we can find a gadget chain for this case, then using *Parcelable* classes is inherently as dangerous as *Serializable*. Thereby, this research question expands on *Bechler's* analysis of deserialization vulnerabilities outside of *Java* native serialization, which did not cover the *Parcel API* [10].

We run *AndroChain* over the entire dataset outlined in Section 4.1 and configure `Parcel.readSerializable()`, and `Bundle.readSerializable()` as sink methods. From our search, we find three candidate gadget chains that trigger *Java* native serialization through a *Parcelable* (see Listing 13). Two of these originate from the *Android Framework SDK* and are hence available via *Zygote* by default. The third gadget chain stems from the *Google Play Services Cast SDK* [51].

```
(1) android.net.wifi.hotspot2.PasspointConfiguration$1
    -> android.os.Bundle.getSerializable()
(2) android.net.wifi.hotspot2.OsuProvider$1
    -> android.os.Bundle.getSerializable()
(3) com.google.android.gms.cast.tv.media.zzab
    -> [...].tv.media.StoreSessionRequestData.<init>
        -> android.os.Bundle.getSerializable()
```

**Listing 13:** *Parcelable* creators (`createFromParcel()`) enabling a *Java* native deserialization entry point.

In all three instances, the call to `getSerializable()` is trivially reachable through the `createFromParcel()` method. For example, the parcelable class `StoreSessionRequestData` (see Listing 14) deserializes an arbitrary `Serializable` from the *Bundle* `b2` in its constructor (line 5). Then, the constructor is reached through the defined `Parcelable.Creator` `zzab` (line 2,7) which unparcels `b2` (line 17) and passes it to initialization in line 19.

|  | **java.lang.Object** | | | **java.util.Map** | | **java.util.Comparator** | **java.util.Comparable** |
|---|---|---|---|---|---|---|---|
|  | *hashCode* | *toString* | *equals* | *put* | *get* | *compare* | *compareTo* |
| Reachability | ◐ | ● | ◐ | × | ◐ | ◐ | × |
| *Ysoserial* chains | 9 | 2 | 2 | 1 | 4 | 5 | 0 |

**Table 4:** Reachable *trampoline gadgets*, and occurrence count in *Ysoserial*. Android - ○, JCL - ●, both - ◐, neither - ×

```
1  class StoreSessionRequestData {
2    Parcelable.Creator CREATOR = new zzab();
3    StoreSessionRequestData(Bundle b1, Bundle b2) {
4      Serializable var4 = b2
5        .getSerializable("targetDeviceCapabilities");
6  }}
7  class zzab implements Parcelable.Creator {
8  public Object createFromParcel(Parcel p){
9    Bundle b1 = null;
10   Bundle b2 = null;
11   while (p.dataPosition() < validateHeader) {
12     int readHeader = SafeParcelReader.readHeader(p);
13     switch (SafeParcelReader.getFieldId(readHeader)) {
14     case 2: // reconstruct Bundle b1
15       break;
16     case 3:
17       b2 = SafeParcelReader.createBundle(p, readHeader);
18       break;}}
19   return new StoreSessionRequestData(b1, b2);}}
```

**Listing 14:** Decompiled `StoreSessionRequestData` in *Google Play Services Cast SDK*. The `Bundle` bundle2, read on line 17 is passed to the constructor (line 19) which then triggers arbitrary deserialization (line 5).

To achieve arbitrary deserialization, we modify the `writeToParcel()` method of `StoreSessionRequestData`. I.e., we copy the class to the sending app and rewrite the critical lines of code responsible for serialization (see Listing 15, lines 3, 4). When wrapping the modified *Parcelable* into an *Intent* and calling `startActivity()`, *Android* will parcel according to the custom `writeToParcel()` and then unparcel on the recipient's side with the original `createFromParcel()`, containing the type-unsafe deserialization entry point. We provide a PoC for this scenario in Section 8.

```
1  public void writeToParcel(Parcel dest, int flags) {
2    Bundle a = new Bundle();
3    this.zza.putSerializable(
4      "targetDeviceCapabilities", payload);
5    Bundle b = new Bundle();
6    int var6 = SafeParcelWriter.beginObjectHeader(dest);
7    SafeParcelWriter.writeBundle(dest, 2, a, false);
8    SafeParcelWriter.writeBundle(dest, 3, b, false);
9    SafeParcelWriter.finishObjectHeader(dest, var6);
```

**Listing 15:** Modified `writeToParcel` in mocked *Google Play Services Cast* `StoreSessionRequestData`.

Triggering arbitrary deserialization through the entry points in the *Android Framework SDK* ((1) and (2) in Table 13) is conceptually the same, albeit modifying `writeToParcel()` requires more effort due to naming conflicts with the shared classes from *Zygote's boot.art* [17]. We provide the entry points in Appendix C for reference. On a side note, arbitrary deparcelization through an unsafe deserialization entry point is also achievable (see Appendix D). This is interesting, since if a vulnerability existed in the *Parcel API* it would indirectly affect deserialization.

### RQ3: Is it possible to find a security-sensitive gadget chain in the Android ecosystem?

We now search for gadget chains in the entire dataset outlined in Section 4.1. That is, on top of *Android's libcore* and *framework SDK*, we include all 1 200 dependencies collected from the *Google Maven* repository and from F-Droid. The initial configuration uses an empty ignore list and the entire sink method dataset gathered from Section 4.1.2. We run *AndroChain* iteratively, manually refining the ignore and sink list according to identified false positives. Section 3.1 explains why adding to the ignore list is crucial. In essence, we avoid path explosion by visiting each gadget only once, which can, filter out some variations leading to the same sink method. Therefore, when we detect a variant and categorize it as false positive, we must add it to the ignore list. Rerunning the tool at this point detects further gadget chains. As mentioned in Section 4.1.2, we may have to reduce the *Android*-specific sink list from *SuSi* since it is outdated and not completely accurate due to the machine learning model features [6].

We run three campaigns for different entry points. That is (1) *Java* native serialization entry points (`readObject()`, `readResolve()`, and `readObjectNoData()`), (2) dynamic proxies (serializable subtypes of `InvocationHandler.invoke()`), and (3) any necessary follow-ups for reflective call sites. Thereby, campaigns (2) and (3) benefit from the updated ignore- and *SuSi* sink-list in the previous campaign(s). The results of the three campaigns are presented in Table 5.

From the initial run, we find gadget chains leading to an arbitrary invocation of `Class.forName()` in libraries using *Google Protocol Buffers* (e.g., Appendix E). This shows it is generally possible to call the static initializer (`<clinit>`) of any (including non-serializable) class within the victim application. We add this as an entry point to the third campaign.

| Run | GC | Ignore List | Duration | Method Coverage |
|---|---|---|---|---|
| **(1)** *initial* | 3 798 | 0 | 591.15s | 95 046 |
| **(1)** *1* | 26 | 51 | 49.50s | 20 609 |
| **(1)** *2* | 9 | 66 | 43.20s | 19 866 |
| **(1)** *3* | 4 | 71 | 45.76s | 19 856 |
| **(1)** *4* | 2 | 77 | 50.36s | 19 837 |
| **(1)** *5* | 0 | 78 | 42.42s | 15 699 |
| **(2)** *Proxy* | 2 | 78 | 45.61s | 14 094 |
| **(3)** *Clinit* | 3 003 | 78 | 1 052,29s | 89 114 |

Table 5: Statistics for the individual *AndroChain* runs by campaign (1) deserialization methods, (2) *InvocationHandler.invoke()*, (3) taintable reflective call-sites found in (1).

**Reachability of Kotlin Lambda Functions**

In the inital run, we notice a pattern that many chains are related to the potential reachability of arbitrary *Kotlin* lambda function invocation. Unlike *Java* native lambda functions, which get compiled into the *invokedynamic* bootstrap section of a class file, *Kotlin*-defined lambda functions in source code are compiled into separate (anonymous) *Java* classes within bytecode (as subtypes of `Function<R>`). It so happens that many of the concrete *Kotlin* lambda function implementations are serializable (for instance, subtypes of `SuspendLambda`). Therefore, to move forward, we first need to confirm whether invocation of arbitrary *Kotlin* lambda functions is possible. If this is not the case, we can eliminate all gadget chain findings relying on this (trampoline) gadget for this and further runs.

Therefore, we rerun *AndroChain* with `Function<R>` subtypes defined as sinks. We find seven gadget chains leading to invocations of `Function0` and `Function2` (see Appendix F). All of these turned out to be false positives. Three are related to *kotlin-stdlib*'s `Lazy` values and their respective subtypes `SafePublicationLazyImpl`, `SynchronizedLazyImpl`, and `UnsafeLazyImpl`. Here, there appears to be a gadget chain `toString() -> getValue() -> Function0.invoke`, which, however, fails due to unsatisfiability of a conditional statement (see Appendix G). Then, for the remaining four chains, there is no arbitrary *Kotlin* lambda function invocation to begin with since a specific implementation is passed to the next gadget as a parameter. Thus, we disregard *Kotlin* lambda functions for further runs.

**Runs 1-5**

We create an ignore list from the false positives in the initial run and the *Kotlin* lambda function run. As a result, *AndroChain* now only reports 26 remaining gadget chains. This significant reduction of gadget chains (from 3 787) with only 51 ignore list entries shows that our strategy to iteratively and manually remove false positives is manageable. At this point, we run *AndroChain* four more times, decreasing the number of chains to 9, 4, 2, and 0, respectively. Overall, **none of the gadget chains found in these runs has a security impact**. For the complete ignore list comprising 78 false positive gadgets, we refer to our artifact (see Section 8).

We highlight two gadget chains we manually verified as false positives. Both instances demonstrate how small the changes are to activate a currently dormant gadget chain. For one, we found gadget chain candidates in the *RxJava* library leading to closing an arbitrary `AutoCloseable`. These, however, cannot be weaponized since *Android* permission-protected `AutoCloseables` like `BluetoothSocket` are not serializable. The second gadget chain, also in *RxJava*, would lead to running an arbitrary task through calling `next()` on an internally defined iterator[3]. Thus, the question becomes whether a call to a generic `Iterator.next()` exists (i.e., whether this is a *trampoline gadget*). Iterating over a *Collection* is a common pattern in native deserialization which leads to an abundance of calls to `Iterator.next()`. These calls are likely false positives since *Collections* use a specific iterator implementation. To minimize the search space, we search for serializable *Collection* subtypes with a declared generic `Iterator` property and found no such class. I.e., there currently is no suitable gadget in which an attacker could place a payload relying on the aforementioned specific *RxJava* `Iterator`. Both the `Iterator`- and the `AutoCloseable`-based gadget chain, have the potential of becoming a liability in the future. All it takes is adding a gadget that makes a trampoline reachable or adding the *Serializable* interface.

**Other Entrypoints (Campaigns 2 and 3)**

We now consider dynamic proxies as entry points. That is, given an interface method being invoked during deserialization, the very method call could be proxied by a serializable `InvocationHandler`. Thus, we run *AndroChain* with `InvocationHandler.invoke()` as an entry point, and the accumulated ignore list. The result comprises two false positives related to an internally used `OutputStream`. This small amount makes sense because *Android's* `AnnotationFactory` is the only serializable `InvocationHandler` in the dataset, and the ignore list already captures false positives related to its `equals()` method.

Finally, due to the finding in campaign (1) of arbitrary static class initialization invocation being possible, we rerun *AndroChain* with the `<clinit>` signature as an entry point. Thereby, we (a) no longer check whether classes are serializable since the (sub-)gadget chain now originates from the *ClassLoader*, (b) use only VTA for polymorphic call sites, and (c) consider all parameters within gadget chains to be untainted. The modifications (b) and (c) are because `<clinit>` itself is parameterless and if any static class properties are set, then from literals within this method [63]. Thus we, only need to check whether any of the reachable sink methods is security-critical (even if untainted) which makes analyzing the resulting 3 003 gadget chains manageable. We collect 105 distinct reachable sink methods, of which all inherited classes of *OutputStream*, logging configuration-related sinks, file system operations, and reflective method invocations are benign. There are two calls to `Runtime.exec()`, calling the binaries /sbin/ldconfig [61] and /system/bin/getprop [50]. Both have no further security implication given there are no vulnerabilities in those binaries.

---
[3] `BlockingFlowableIterable$BlockingFlowableIterator.next`

## 5 Discussion

The bottom line of this work is that while insecure deserialization entry points (through *Intents*) are abundant, we could find no compelling evidence that deserialization gadget chains are a pathological problem in *Android*. In the light of previous research [26]–[28], [36], [40] continuously finding more gadget chains in the *Java* libraries part of *Ysoserial*, this result is surprising. After all, the *Ysoserial* dataset consists of 41, whereas the analysis domain of our study comprises 1 200 dependencies. In addition, we used a much wider sink method dataset, compiled from all the sink lists of previous research and *Android*-specific sinks. Under these circumstances, we anticipated finding at least some gadget chains. This not being the case raises questions about (a) the rigor of our study and, if passing scrutiny, (b) its implications for *Android* and further research in *Java* deserialization vulnerabilities.

### 5.1 Methodology and Rigor

Our approach is distinct from previous work in that it puts a strong emphasis on soundness and efficiency. As such, our gadget chain detector, *AndroChain*, is specifically designed for the analysis of large datasets with an unknown ground truth. We avoid path explosion through an iterative worklist algorithm that visits call sites only once. This implies that at no point is the entire call graph loaded in memory, and consequently it also misses gadget chains that use the same gadget as another chain. By adding false positive call edges to a continuously evolving ignore list, we can ensure that these gadget chains are found in future runs. Evidently, our process requires an experienced *Java* researcher to manually inspect gadget chain findings both to determine false positives and identify recurring gadgets across the result. The latter endeavor drastically minimizes the number (3 798) of gadget chains that require inspection, resulting in a 78-entry ignore list. We argue this effort to be sustainable for large datasets, albeit future work could improve automation.

Focusing on soundness proved to be beneficial for detecting deserialization patterns that were previously invisible. For instance, the gadgets in *Jackson*'s `PrivateMaxEntriesMap` [65] turn out to be a false positive due to the usage of a *SerializationProxy* (see Listing 16). Specifically, gadget chains were relying on invoking arbitrary `Runnable` objects from the `writeBuffer` (line 2). Yet, ordinary deserialization via `readObject()` is denied (line 4). Instead, a *SerializationProxy*'s `readResolve()` method (line 10) reconstructs the `PrivateMaxEntriesMap`. Manual investigation was necessary to determine how the object is reconstructed and whether thereby the `writeBuffer` can be tainted. Similarly, we investigated how *transient* fields are restored in *Android*'s `AnnotationFactory`. Even though both instances are false positives, these specific deserialization routines could have gone either way.

```
1  public final class PrivateMaxEntriesMap<K, V> {
2    final Queue<Runnable> writeBuffer;
3    private void readObject(ObjectInputStream stream) {
4      throw new InvalidObjectException("Proxy required");
5    }
6    static final class SerializationProxy<K, V>
7    implements Serializable {
8      final Map<K, V> data;
9      final long capacity;
10     Object readResolve() {
11       PrivateMaxEntriesMap<> map = new Builder<K, V>()
12         .maximumCapacity(capacity).build();
13       map.putAll(data);
14       return map;
15  }}}
```

**Listing 16:** In *Jackson-Databind*'s `PrivateMaxEntriesMap` the queue `writeBuffer` cannot be tainted due to the delegation of deserialization to a `SerializationProxy`.

In Section 3.2, we benchmark *AndroChain*'s soundness in comparison to purely static gadget chain detection tools [14], [28], [40]. We refrain from taking hybrid tools (e.g., *Crystallizer* [32] and *JDD* [36]) for this comparison since their contribution is mainly increasing precision by incorporating a fuzzer on top of the static analysis. A recent study [47], however, proves how this practice compromises soundness. Specifically, according to this study, the purely static state-of-the-art tool *Tabby* proves most sound on a synthetic benchmark. The results of the tool evaluation show that *AndroChain* keeps up with *Tabby*'s soundness, increases performance, and still produces a meaningful output that can be manually analyzed in reasonable time.

### 5.2 Implications and Future Work

The inability to find any gadget chain in a large-scale dataset of *Android* dependencies does not imply that insecure deserialization is a negligible attack vector. Indeed, minor modifications of selected classes would enable some gadget chains (e.g., making the *AutoCloseable* class serializable). According to [31], such minor modifications are the main cause for gadget chains surfacing. Thus, for *Android* targeting API levels of 33 and above, we strongly advocate the usage of the newly added type-safe *Intent* deserialization and deparcelization methods [57]. However, from our results, apps running on older *Android* versions do not seem to be at risk of insecure deserialization gadget chains. We make this statement with necessary precaution, as our analysis is limited by the challenges of *Java* static analysis (e.g., not being able to analyze native libraries). Furthermore, our study considers dependencies in their latest version. This gives no indication of whether gadget chains existed in earlier versions. Doing so would require an effort comparable to [39] on a substantially larger dataset and myriad dependency combinations, which is outside of this work's scope.

Regardless, our analysis evokes the question of whether *Java* deserialization gadget chains are as widespread as previous research indicates. We observe a pattern where first a tool is tested on the *Ysoserial* dataset, and then hand-picked *Java* programs are analyzed by the tool, which results in gadget chain findings. For instance, *JDD*, states "having found 127 zero-day gadget chains in six real-world *Java* applications" [36]. *Tabby* finds five new gadget chains in three middleware apps [28]. *Crystallizer* identifies two vulnerabilities in *Apache Kafka* and *Pulsar*, respectively [32]. All three works show that, indeed, deserialization gadget chains continue to be a liability to current software. However, the reason why certain *Java* applications are analyzed remains unclear. To a greater extent, this could indicate that for the publications, a much larger dataset of applications was analyzed, yet only the successful results were reported. This definitely demonstrates the tools' capability, but it gives the impression that *Java* deserialization gadget chains are widespread. It thereby contradicts the incapability of finding any security-critical gadget chains in a large-scale *Android* dependency dataset.

Our results could be coincidental to the dataset of *Android* dependencies and the *Android* framework SDK. Even though we could find no evidence that *Android* is by design more robust against deserialization gadget chains (i.e., Section 2.4, RQ1, RQ2, and [57]), this bias deserves consideration. Therefore, we suggest replicating our study on a different large-scale dataset of *Java* dependencies and applications. Furthermore, noting that minor modifications in *Java* source code can weaponize currently dormant gadget chains, we plan to investigate this aspect in regard to supply chain attacks.

## 6 Related Work

*Java* deserialization gadget chain detection tools have been developed in [11], [14], [21], [22], [24]–[29], [32], [36], [40], [41], [43]. The main difference to *AndroChain* is that it is designed to analyze a large-scale dataset of JAR files efficiently, and the usage of an iterative worklist algorithm combined with an evolving ignore list. As such, the premise of our study is also distinct. Where all the mentioned publications' main contribution is an improvement to gadget chain detection methodology, our contribution lies in the analysis of *Android* and building a tool catered for this task.

*Peles et al.* made use of *Android Intent* deserialization in [8] to exploit native pointers declared in *Java* application code. *Graux et al.* expanded this analysis to a larger scale [20]. Both works, however, investigated native pointers not declared as *transient* instead of gadget chains. There are also two PoC applications online [42], [44] that port a gadget chain from *Ysoserial* to an *Android* application. While this proves that gadget chains are a functional attack vector in *Android*, it conducts no analysis on finding gadget chains in *Android* and common dependencies.

Adjacent to our preliminary analysis in Section 2.4 of insecure deserialization entry points in *Android*, one finds *ObjectMap* [16], a scanner for deserialization entry points in *Java*-based web applications. Where *ObjectMap* infers entry points through web responses, our analysis is a white-box setting from given APK files. Moreover, our main contribution is searching for gadget chains, whereas *ObjectMap* tests web endpoints with preexisting payloads from *Ysoserial*.

*FlowDroid* [4] and *TaintDroid* [5] are static analysis frameworks for *Android* applications. These tools model the app lifecycle or system-wide taint-tracking, neither of which apply to our studies' domain. Finally, there is a range of foundational research on *Java* deserialization gadget chains: from concept [10], [23], [37], their evolution over dependency versions [31], [39], to usage for memory exhaustion attacks [12].

## 7 Conclusion

In this work, we assessed the attack surface of *Java* deserialization gadget chains in *Android*. We found that unsafe deserialization entry points are widely available either directly in *Android* system apps or other widely distributed apps. From there we created a large-scale dataset of *Android* app dependencies and a comprehensive list of security-sensitive sink methods. Also, we built *AndroChain* - a novel gadget chain detection tool that has the capacity to efficiently operate on these large datasets. We verified its performance against other static analysis based gadget chain detectors and found it to achieve at least similarly sound results.

We then showed that the *Android* framework SDK includes most of the *trampoline gadgets* available through the *Java Class Library*. Furthermore, by demonstrating *Java* deserialization through the *Android Parcel API*, this mechanism proved not to be inherently more secure than the *Java Serializable* API. All experimentation results up to this point gave no evidence that *Android* has a better security posture against deserialization gadget chains than other *Java* artifacts.

Nevertheless, we found no security-critical gadget chains from analyzing the large-scale (1 200) *Android* dependency dataset in combination with a comprehensive (4 616) sink method list. This result challenges previous research indications that *Java* deserialization gadget chains are a widespread problem. Consequently, we propose directing further research not only to the improvement of gadget chain detection methodology but also to analyzing it as a supply-chain attack vector and a more rigorous analysis of the entire *Java* platform.

# 8 Open Science

We share our tool *AndroChain* with Link 1, experimentation data, and scripts for dataset generation with Link 2. Both repositories contain a README file with instructions on using/reproducing. For reproducibility, we also share the AOSP build artifacts, APK files used in Section 2.4 and the dependency dataset used in Section 4 with Link 3. That is because the download pipelines shared in Link 2 will likely produce different datasets in the future.

1. `https://zenodo.org/records/14608101?preview=1&token=eyJhbGciOiJIUzUxMiIsImlhdCI6MTczNjIzNzc5NSwiZXhwIjoxNzQ4NzM1OTk5fQ.eyJpZCI6Ijc0OTI0NjQ5LTczYmEtNGJhZS1hMzdiLTVlMzk2NDM0MTY5YSIsImRhdGEiOnt9LCJyYW5kb20iOiJiZjA3OGNlODg0OTI1OWQwYTQ2NDllOTIyNTJmZDNhYSJ9.Wh7DgTFsl3QUoLkA5QdE-NSCJnKGGQ7xFyknNl_9pyahChsSi3oqI9t5jPIEsvrgVnkx0RwOfJQO7XjKi1wuMQ`

2. `https://zenodo.org/records/14614035?preview=1&token=eyJhbGciOiJIUzUxMiIsImlhdCI6MTczNjMxNzIwMiwiZXhwIjoxNzQ4NzM1OTk5fQ.eyJpZCI6ImQ2NmFmZWRlLWRiMmItNGM1ZS1hM2RiLWY1NzE3ZTIyOGJkNyIsImRhdGEiOnt9LCJyYW5kb20iOiIzMThhNzlhOWE5YWY2ZTYwNzA5ZjUyNDcyZDY2ZjlmOCJ9.xTa6Yq0CFOlAXI2F0u48Vk226zVZKipttYLjZPZyZ9xa2O1D8ksKgLV5wr0GAw8NQqoa4cKpB96KQj29DoWgPw`

3. `https://zenodo.org/records/14607806?preview=1&token=eyJhbGciOiJIUzUxMiIsImlhdCI6MTczNjIzNDg1NywiZXhwIjoxNzQ4NzM1OTk5fQ.eyJpZCI6IjEzOThkYzRhLWFmNDEtNGM4Yi1hZjg2LThiYmIxODI2NDA2NSIsImRhdGEiOnt9LCJyYW5kb20iOiI2MmM3OGUwOWI0YzhmNjZhOGEyOTQ2YjFmNiJ9.FgbCQ9PiL30l31YOVmOEh4uPCCbAAl3PvDUnvKmwuU1SKihzVqGdQxU4tAOvTQd3iCVH3DSfeIfQErdfWjlqbQ`

# 9 Ethics Considerations

The data processed in this publication involves no human or human-relatable data. The datasets we used in our analysis are open source, so no permission was needed prior to performing this study. In terms of security information disclosure, there are no concrete vulnerabilities or exploits discussed in this work and its artifacts. Thus, there was no necessity to abide by any responsible vulnerability disclosure guidelines.

Overall, this publication puts no individuals or organizations at substantial risk. The attack vectors described are mostly known from prior work, with the exception of showing arbitrary deserialization through the *Android Parcel API*. This can be considered a weakness that may have the potential to turn into a vulnerability through a gadget chain in the future. Yet again, we found no concrete gadget chains with a security implication. In line with related work in this research field and the *Open Science Policy*, we release our tool *AndroChain* to the community. As a vulnerability detection tool, *AndroChain* can be used both by malicious actors to find and exploit currently unknown gadget chains, and by software developers to harden their applications. We are aware of this trade-off [66] and believe it is in the best interest of our society to not withhold the information and availability of such a tool from the public.

# Appendix A   Java Sink Method Lists

- **Serianalyzer** https://github.com/mbechler/serianalyzer/blob/master/src/main/java/serianalyzer/SerianalyzerConfig.java#L570
- **GadgetInspector** https://github.com/JackOfMostTrades/gadgetinspector/blob/master/src/main/java/gadgetinspector/GadgetChainDiscovery.java#L236
- **Tabby** https://github.com/wh1t3p1g/tabby/blob/master/rules/sinks.json
- **JDD** https://github.com/fdu-sec/JDD/tree/main/src/jdd/rules/sinks
- **SerdeSniffer** https://github.com/SerdeSniffer/SerdeSniffer/blob/main/tools/custom-rules/summary/config/Configuration.dl

# Appendix B   Gleipner Benchmark Results

| Category | Tabby | | AndroChain | |
|---|---|---|---|---|
| | TP | FP | TP | FP |
| **Depth** | 20 | 0 | 20 | 0 |
| **Polymorphism** | 20 | 0 | 20 | 0 |
| **Multipath** | 10 | 0 | 1 | 0 |
| **Payload Construction** | 9 | 6 | 9 | 6 |
| **Serialization API** | 4 | 0 | 3 | 0 |
| **Serialization Keywords** | 2 | 1 | 2 | 1 |
| **Method.invoke()** | 1 | 1 | 1 | 1 |
| **Class Initialization** | 0 | 0 | 0 | 0 |
| **Classloading** | 1 | 1 | 1 | 1 |
| **Constructor** | 1 | 1 | 1 | 1 |
| **Dynamic Proxies** | 1 | 1 | 3 | 2 |
| **Reflection Exceptions** | 7 | 7 | 7 | 7 |
| **Meta Objects** | 10 | 10 | 10 | 10 |
| **Runtime Exceptions** | 7 | 7 | 7 | 7 |
| **JNI Methods** | 0 | 0 | 0 | 0 |
| **Invoke Dynamic** | 1 | 1 | 1 | 1 |
| **Ysoserial** | 7 | 0 | 7 | 0 |
| **Total TP** | 101 | | 93 | |

**Table 6:** Results of running *AndroChain* on the *Gleipner* benchmark [47]. *Tabby*'s results are added for reference as it achieved the most sound results on the benchmark in [47]. True positives (TP) in green indicate that all true positives in this category were found. Conversely, false positives (FP) in red indicate that the tool is imprecise since false positives are reported. There are three categories where *Tabby* and *AndroChain* perform differently. *AndroChain* finds only one **Multipath** gadget chain because of the ignore list heuristic; if we rerun *AndroChain* nine additional times with an evolving ignore list, we find the remaining nine gadget chains. In **Serialization API**, *Tabby* was additionally tested with the `Externalizable.readExternal()` entry point. This deserialization method has no relevance to *Android* and can trivially be added to the tool if needed. By searching from *InvocationHandler.invoke()* as an entry point, *AndroChain* finds all **dynamic proxies**. It should also be mentioned that the seven **Ysoserial replica** gadget chains found by *Tabby* and *AndroChain* are not identical. Then again, in [47] an *APOC* query was used instead of *tabby-path-finder* to query gadget chains from the *Neo4J* database. Leaving that aside, the only negative difference in terms of soundness lies within the **Multipath** and **Serialization API** categories. Using an ignore list over multiple runs alleviates the former, and the latter is an irrelevant entry point for our study.

## Appendix C  Arbitrary Deserialization in *Android Framework SDK*

```
1  public void writeToParcel(Parcel dest, int flags) {
2    Bundle bundle = new Bundle();
3    // omitted parcel writes
4    bundle.putSerializable("serviceFriendlyNames",
5      (HashMap<String, String>) mServiceFriendlyNames);
6    // omitted parcel writes
7  }
8
9  PasspointConfiguration createFromParcel(Parcel in) {
10   PasspointConfiguration config =
11     new PasspointConfiguration();
12   // omitted parcel reads
13   Bundle bundle = in.readBundle();
14   Map<String, String> friendlyNamesMap = (HashMap)
15     bundle.getSerializable("serviceFriendlyNames");
16   // omitted parcel reads
17   return config;
18 }}
```

**Listing 17:** android.net.wifi.hotspot2.PasspointConfiguration. Modifying lines 4 and 5 in `writeToParcel()` enables arbitrary deserialization through the *Parcel API* in line 15.

```
1  public void writeToParcel(Parcel dest, int flags) {
2    // omitted parcel writes
3    Bundle bundle = new Bundle();
4    bundle.putSerializable("friendlyNameMap",
5      (HashMap<String, String>) mFriendlyNames);
6    dest.writeBundle(bundle);
7  }
8
9  public OsuProvider createFromParcel(Parcel in) {
10   // omitted parcel reads
11   Bundle bundle = in.readBundle();
12   Map<String, String> friendlyNamesMap = (HashMap)
13     bundle.getSerializable("friendlyNameMap");
14   return new OsuProvider(...);
15 }
```

**Listing 18:** android.net.wifi.hotspot2.OsuProvider. Modifying lines 4 and 5 in `writeToParcel()` enables arbitrary deserialization through the *Parcel API* in line 13.

## Appendix D  Arbitrary Deparcelization through *Serializable* in *Google Navigation SDK*

```
1  // class com.google.android.libraries
2  // .navigation.internal.pr.ab
3  private void readObject(ObjectInputStream stream) {
4    stream.defaultReadObject();
5    int readInt = stream.readInt();
6    byte[] bArr = new byte[readInt];
7    stream.readFully(bArr);
8    Class<?> cls = getClass();
9    int length = bArr.length;
10   ClassLoader classLoader = cls.getClassLoader();
11   Parcel obtain = Parcel.obtain();
12   obtain.unmarshall(bArr, 0, length);
13   obtain.setDataPosition(0);
14   Parcelable readParcelable =
15     obtain.readParcelable(classLoader);
16   obtain.recycle();
17   this.f = (Bitmap) readParcelable;
18 }
19
20 private void writeObject(ObjectOutputStream stream){
21   stream.defaultWriteObject();
22   Bitmap bitmap = this.f;
23   Parcel obtain = Parcel.obtain();
24   obtain.writeParcelable(bitmap, 0);
25   byte[] marshall = obtain.marshall();
26   obtain.recycle();
27   stream.writeInt(marshall.length);
28   stream.write(marshall);
29 }
```

**Listing 19:** Internal obfuscated class in *Google Navigation SDK* [53]. One can modify line 22 to an arbitrary *Parcelable* object, which is read on line 15 upon deserialization.

## Appendix E  Arbitrary Call to `<clinit>`

```java
package androidx.datastore.preferences.protobuf;

public abstract class GeneratedMessageLite {
  protected static final class SerializedForm
  implements Serializable {
    private final Class<?> messageClass;
    private final String messageClassName;
    private final byte[] asBytes;

    protected Object readResolve()
    throws ObjectStreamException {
      try {
        Class<?> messageClass =
          resolveMessageClass();
        // remainder omitted, irrelevant
      } catch (Exception e) {
        // ...
      }}

    private Class<?> resolveMessageClass()
    throws ClassNotFoundException {
      return this.messageClass != null ?
        this.messageClass :
        Class.forName(this.messageClassName);
    }
}}
```

**Listing 20:** Arbitrary invocation of `Class.forName()` in *Android Jetpack DataStore preferences core* library.

## Appendix F  Unreachable Kotlin Lambda Function Invocation

| |
|---|
| `kotlin.SafePublicationLazyImpl.toString()` <br> ↪ `kotlin.SafePublicationLazyImpl.getValue()` <br> ↪ `kotlin.jvm.functions.Function0.invoke()` |
| `kotlin.SynchronizedLazyImpl.toString()` <br> ↪ `kotlin.SynchronizedLazyImpl.getValue()` <br> ↪ `kotlin.jvm.functions.Function0.invoke()` |
| `kotlin.UnsafeLazyImpl.toString()` <br> ↪ `kotlin.UnsafeLazyImpl.getValue()` <br> ↪ `kotlin.jvm.functions.Function0.invoke()` |
| `kotlin.coroutines.CombinedContext.toString()` <br> ↪ `kotlin.coroutines.CombinedContext.fold()` <br> ↪ `kotlin.jvm.functions.Function2.invoke()` |
| `kotlin.coroutines.CombinedContext$Serialized.readResolve()` <br> ↪ `kotlin.coroutines.CoroutineContext$DefaultImpls.plus()` <br> ↪ `kotlin.coroutines.CombinedContext.fold()` <br> ↪ `kotlin.jvm.functions.Function2.invoke()` |
| `android.support.wearable.complications.ComplicationData$SerializedForm.readObject()` <br> ↪ `android.support.wearable.complications.ComplicationDataKt.readNullable()` <br> ↪ `kotlin.jvm.functions.Function0.invoke()` |
| `android.support.wearable.complications.ComplicationData$SerializedForm.readObject()` <br> ↪ `android.support.wearable.complications.ComplicationDataKt.readList()` <br> ↪ `kotlin.jvm.functions.Function0.invoke()` |

**Table 7:** False positive gadget chains in *kotlin-stdlib* and *androidx watchface-complication-data* leading to *Kotlin* lambda function invocation.

## Appendix G  Unreachable Kotlin Lambda Function Invocation in *Kotlin* `Lazy`

```
 1  package kotlin;
 2  public final class UnsafeLazyImpl<T>
 3    implements Lazy<T>, Serializable {
 4    private Function0<? extends T> initializer;
 5    private Object _value;
 6
 7    public T getValue() {
 8      if (this._value == UNINITIALIZED_VALUE.INSTANCE) {
 9        Function0<? extends T> function0 =
10          this.initializer;
11        Intrinsics.checkNotNull(function0);
12        this._value = function0.invoke();
13        this.initializer = null;
14      }
15      return (T) this._value;
16    }
17
18    public boolean isInitialized() {
19      return this._value !=
20        UNINITIALIZED_VALUE.INSTANCE;
21    }
22
23    public String toString() {
24      return isInitialized() ?
25        String.valueOf(getValue()) :
26        "Lazy value not initialized yet.";
27    }
28  }
```

**Listing 21:** The call to an arbitrary `Function0` is unfeasible from `toString()` because reaching `getValue()` requires `isInitialized()` to be `true`. However, if that is the case `function0.invoke()` (line 11) is unreachable because `this._value` needs to be uninitialized to satisfy the if-statement.